\documentclass[12pt]{article}
\usepackage{yfonts}
\usepackage{amsmath}
\usepackage{amssymb}
\usepackage{amstext}
\usepackage{amscd}
\usepackage{amsfonts}
\newcommand{\be}{\begin{equation}}
\newcommand{\nd}{\noindent}
\newcommand{\ee}{\end{equation}}
\newcommand{\ben}{\begin{eqnarray}}
\newcommand{\een}{\end{eqnarray}}
\newcommand{\nn}{\nonumber \\}

\title{{\bf The Tsallis-Laplace Transform
}}

\author{ A. Plastino$^1$ and M. C. Rocca$^{1,\,2}$ \\$^1$
Instituto de F\'{\i}sica La Plata - CCT-Conicet\\Universidad
Nacional (UNLP)\\C.C. 727 (1900) La Plata, Argentina\\ $^2$
Departamento de F\'{\i}sica, Fac. de C. Exactas, UNLP}

\date{February 8, 2013}

\def\baselinestretch{2}

\begin{document}

\maketitle
\begin{abstract}

\nd We introduce here the q-Laplace transform as a new weapon in
Tsallis' arsenal, discussing its main properties and analyzing
some examples. The  q-Gaussian instance receives special
consideration. Also, we derive the q-partition function from the
q-Laplace transform.

\nd KEYWORDS: q-Laplace transform, tempered ultradistributions,
complex-plane generalization, one-to-one character.

\end{abstract}

\newpage

\renewcommand{\theequation}{\arabic{section}.\arabic{equation}}

\section{Introduction}
\subsection{The Laplace transform}
\nd The Laplace transform,  introduced by Laplace in his work on
probability theory, is a widely used integral transform with many
applications in physics and engineering. It is a linear operator
acting on  a function $f(t)$ with a real argument $t$) that
transforms it to a function $F(s)$ with complex argument $s$,

\be \label{definit}
\mathcal{L}\{f(t)\}(s)=\int_0^\infty\,dt\,e^{-st}\,f(t);\,\,s\,\,complex.
\ee

\nd This transformation is essentially bijective for the majority
of practical uses; the respective pairs $f(t),\,\,\,L(s)$ are
matched in tables. The Laplace transform has the useful property
that many relationships and operations over the originals $f(t)$
correspond to simpler relationships and operations over the images
$L(s)$. The Laplace transform is related to the Fourier transform
(FT), but whereas the FT transform expresses a function or signal
as a series of modes of vibration (frequencies), the Laplace one
resolves a function into its moments. Customarily, in speaking of
the  Laplace transform without qualification one means the
unilateral or one-sided transform. The Laplace transform can be
alternatively defined as the bilateral, or two-sided one, by
extending the limits of integration to be the entire real axis. If
that is done the common unilateral transform simply becomes a
special case of the bilateral transform where the definition of
the function being transformed is multiplied by the Heaviside step
function.

\nd Recently,  the coupling  of Laplace's transform (LP) with
other methods has become a hot topic \cite{r1,r2}. For example, we
can  mention the Yang-Laplace transform for local fractional
differential equations \cite{r1} and the He-Laplace approach, that
couples He's homotopy perturbation technique with Laplace's
transform \cite{r2}. Here we will link LP with q-statistics
\cite{[1],[2],AP}, but without using the so-called q-calculus.

\subsection{q-Statistical theory}
\nd Nonextensive statistical mechanics (NEXT) \cite{[1],[2],AP},
an extension of the standard Boltzmann-Gibbs (BG) one, is used in
variegated scientific areas. NEXT is based on a nonadditive
(although extensive \cite{[3]})  information measure characterized
by the real index q (with q = 1 recovering the  BG entropy). It
has been employed in diverse scenarios such as cold atoms in
dissipative optical lattices \cite{[4]}, dusty plasmas \cite{[5]},
trapped ions \cite{[6]}, spin-glasses \cite{[7]}, turbulence
\cite{[8]}, self-organized criticality \cite{[9]}, high-energy
experiments at LHC/CMS/CERN \cite{[10]} and RHIC/PHENIX/Brookhaven
\cite{[11]}, low-dimensional dissipative maps \cite{[12]}, finance
\cite{[13]}, galaxies \cite{AP1}, Fokker-Planck equation's
applications \cite{AP2}, etc. A typical NEXT feature is that can
it can be advantageously cast using appropriate q-generalizations
of standard mathematical concepts \cite{borges}. Included are, for
instance, the logarithm and exponential functions, addition and
multiplication, etc.

\subsection{Our aims}

\nd Here we add, by recourse to ultradistributions (see Appendix)
the Laplace transform tool to such armory. It is to be pointed out
that, quite recently, an alternative form of the q-Laplace
transform (qLP) has been advanced by Won Sang Chung
\cite{ginebra}, who uses for that purpose so-called q-sums,
q-differences, q-products, and q-ratios, which renders the
treatment rather abstract.

\nd Our qLP is, instead, based on the ordinary version of the four
elementary arithmetic operations.  More importantly, Wong Sang
Chung's definition does not follow the tenets of  the pioneer
paper by Umarov-Tsallis-Steinberg \cite{umarov} that introduced
the qFourier transform.  In particular, the function to be
Wong-transformed does not contain  a q-exponential argument, a
crucial NEXT-aspect that must be respected so as to maintain the
theory's consistency.

\nd The reader is advised to peruse the Appendix before embarking
into our discussion below.

\section{Laplace transform from Fourier one}

\nd Before dealing with the q-Laplace transform it is convenient
to show that  the ordinary bilateral Laplace transform can be
obtained from the complex Fourier transform. Thus,  let (see
Appendix for details and references)
\begin{itemize}
\item $\Lambda_{\infty}$ be the space of distributions of
exponential type, \item   ${\bf {\cal U}}$ or ${\bf {\cal U}_I}$
the space of tempered ultradistributions,  \item  ${\cal F}$ the
Fourier transform connecting them, and \item $H$ the Heaviside
step function.
\end{itemize}

\nd We have
\begin{equation}
\label{ep6.1} {\cal
F}:{\Lambda}_{\infty}\longrightarrow\boldsymbol{{\cal U}},
\end{equation}
 reading ($\Im(k)$ is the imaginary part of $k$ and $\Re(k)$ its real one)
\begin{equation}
\label{ep6.2} {\cal F}(k)=H[\Im(k)]\int\limits_0^{\infty}
f(x)e^{ikx}\;dx- H[-\Im(k)]\int\limits_{-\infty}^0
f(x)e^{ikx}\;dx.
\end{equation}
The associated inversion formula is
\begin{equation}
\label{ep6.3} f(x)=\frac {1} {2\pi} \oint\limits_{\Gamma_F} {\cal
F}(k)e^{-ikx}\;dk,
\end{equation}
where the contour $\Gamma_F$ surrounds all singularities  of
${\cal F}(k)$ and runs parallel to the real axis from $-\infty$ to
$\infty$ above it, and from $\infty$ to $-\infty$ below. Then,
making the change of variables $ik=-p$, the bilateral Laplace
transform is obtained (see \cite{tp1})
\begin{equation}
\label{ep6.4}
{\cal L}:{\Lambda}_{\infty}\longrightarrow\boldsymbol{{\cal U}_I}
\end{equation}
and given by
\begin{equation}
\label{ep6.5} {\cal L}(p)=H[\Re(p)]\int\limits_0^{\infty}
f(x)e^{-px}\;dx- H[-\Re(p)]\int\limits_{-\infty}^0
f(x)e^{-px}\;dx.
\end{equation}
Let us insist: ${\bf {\cal U}_I}$ is the space of tempered
ultradistributions and  we have made the change of variables
$p=-ik$.  Now, the  corresponding inversion formula is:
\begin{equation}
\label{ep6.6} f(x)=\frac {1} {2\pi i} \oint\limits_{\Gamma_L}
{\cal L}(p)e^{px}\;dp,
\end{equation}
where the contour $\Gamma_L$ surrounds all singularities  of
${\cal L}(p)$ and runs parallel to the imaginary axis from
$-i\infty$ to $i\infty$ to the right of it, and from $i\infty$ to
$-i\infty$ to the left. If we consider distributions of
exponential type $f(x)$ such that $f(x)=0$ for $x<0$, we obtain
the single Laplace transform
\begin{equation}
\label{ep6.7} {\cal L}(p)=H[\Re(p)]\int\limits_0^{\infty}
f(x)e^{-px}\;dx,
\end{equation}
whose inversion formula is
\begin{equation}
\label{ep6.8} f(x)=\frac {1} {2\pi i} \int\limits_{\Gamma_{L_+}}
{\cal L}(p)e^{px}\;dp,
\end{equation}
where $\Gamma_L^+$ is the right hand side of the path $\Gamma_L$.

\section{q-Laplace transform}

\nd Let $\Omega$ stand  the space of functions of the real
variable $x$ that are parameterized by a  real parameters $q$. We
have defined in  (\cite{tp4} - \cite{tp5}) the q-Fourier transform
\begin{equation}
\label{ep6.9}
F:\Omega\longrightarrow\boldsymbol{{\cal U}}
\end{equation}
as
\[F(f)(k,q)=F(k,q)=[H(q-1)-H(q-2)]\times \]
\[\left\{H[\Im(k)]\int\limits_0^{\infty} f(x)\{1+i(1-q)kx[f(x)]^{(q-1)}\}^{\frac {1}
{1-q}},
\;dx -\right.\]
\begin{equation}
\label{ep6.10}
\left. H[-\Im(k)]\int\limits_{-\infty}^0 f(x)
\{1+i(1-q)kx[f(x)]^{(q-1)}\}^{\frac {1} {1-q}} \;dx\right\}.
\end{equation}
and its inverse transform as
\begin{equation}
\label{ep6.11}
f(x)=\frac {1}
{2\pi}\oint\limits_{\Gamma_F}\left[\lim_{\epsilon\rightarrow 0^+}
\int\limits_1^2 F(k,q)\delta (q-1-\epsilon)\;dq\right]
e^{-ikx}\;dk.
\end{equation}
 Let $\Omega_I$ be the space of functions of the real variable
$x$
\begin{equation}
\label{ep6.12}
\Omega_I=\{f(x)/f(x)\in{\Omega}_I^+\cap{\Omega}_I^-\},
\end{equation}
where
\[{\Omega}_I^+=\left\{f(x)/f(x)\{1-(1-q)px[f(x)]^{(q-1)}\}^{\frac {1} {1-q}}\in
{\cal L}^1[\mathbb{R}^+];\right.\]
\[f(x)\geq 0; \left|f(x)\right|\leq |x|^sg(x)e^{ax};s,a\in\mathbb{R}^+;
p\in\mathbb{Z};\Re(p)\geq 0\]
\begin{equation}
\label{ep6.13} \left.1\leq q<2\right\},
\end{equation}
and
\[{\Omega}_I^-=\left\{f(x)/f(x)\{1-(1-q)px[f(x)]^{(q-1)}\}^{\frac {1} {1-q}}\in
{\cal L}^1[\mathbb{R}^-];\right.\]
\[f(x)\geq 0; \left|f(x)\right|\leq |x|^sg(x)e^{ax};s,a\in\mathbb{R}^+;
p\in\mathbb{Z};\Re(p)\leq 0\]
\begin{equation}
\label{ep6.14} \left.1\leq q<2\right\},
\end{equation}
Making again the change $ik=-p$  we immediately obtain the
bilateral q-Laplace transform $L$
\begin{equation}
\label{ep6.15} L:\Omega_I\longrightarrow\boldsymbol{{\cal U}_I},
\end{equation}
as
\[L(f)(p,q)=L(p,q)=[H(q-1)-H(q-2)]\times \]
\[\left\{H[\Re(p)]\int\limits_0^{\infty} f(x)\{1-(1-q)px[f(x)]^{(q-1)}\}^{\frac {1}
{1-q}},
\;dx -\right.\]
\begin{equation}
\label{ep6.16}
\left. H[-\Re(p)]\int\limits_{-\infty}^0 f(x)
\{1-(1-q)px[f(x)]^{(q-1)}\}^{\frac {1} {1-q}} \;dx\right\}.
\end{equation}
The corresponding inversion formula is  easily found from
(\ref{ep6.11})
\begin{equation}
\label{ep6.17}
f(x)=\frac {1}
{2\pi i}\oint\limits_{\Gamma_L}\left[\lim_{\epsilon\rightarrow 0^+}
\int\limits_1^2 L(p,q)\delta (q-1-\epsilon)\;dq\right]
e^{px}\;dk.
\end{equation}
If we consider $f\in\Omega_I$ such that $f(x)=0$ for $x<0$ we
obtain the unilateral q-Laplace transform
\[L(p,q)=[H(q-1)-H(q-2)]\times \]
\begin{equation}
\label{ep6.18} H[\Re(p)]\int\limits_0^{\infty}
f(x)\{1-(1-q)px[f(x)]^{(q-1)}\}^{\frac {1} {1-q}}\;dx,
\end{equation}
and its inversion formula
\begin{equation}
\label{ep6.19}
f(x)=\frac {1}
{2\pi i}\int\limits_{\Gamma_{L_+}}\left[\lim_{\epsilon\rightarrow 0^+}
\int\limits_1^2 L(p,q)\delta (q-1-\epsilon)\;dq\right]
e^{px}\;dk.
\end{equation}
We consider now functions $f_{q^{'}}\in\Omega_I$ depending on the parameter
$q^{'}$ with $1\leq q^{'}<2$. We can define the singular q-Laplace
transform
\begin{equation}
\label{ep6.20} L_R:\Omega_I\longrightarrow\boldsymbol{{\cal U}_I},
\end{equation}
as
\begin{equation}
\label{ep6.21} L_R(f_q^{'})(p,q^{'})=L_R(p,q^{'})=
\lim_{q\rightarrow q^{'}}L(f_{q^{'}})(p,q)=
L(f_{q^{'}})(p,q)\mid_{q=q^{'}},
\end{equation}
As is the case of the $F_T$ q-Fourier transform (see \cite{tp4}), $L_R$ is NOT one to one. To deal with such an issue
we consider the set ${\Lambda}_{If_{q^{'}}}$ given by
\begin{equation}
\label{ep6.22} {\Lambda}_{If_{q^{'}}}=\left\{g_{q^{'}}\in\Omega_I/
L_R(g_{q^{'}})(k)=L_R(f_{q^{'}}(k)\right\},
\end{equation}
and
\begin{equation}
\label{ep6.23}
\Lambda_I=\left\{{\Lambda}_{If_{q^{'}}}/f_{q^{'}}\in\Omega_I\right\}.
\end{equation}
Introducing  the equivalence relation
\begin{equation}
\label{ep6.24} g_{q^{'}}(x)\sim f_{q^{'}}(x)\Longleftrightarrow
g_{q^{'}}\in\Lambda_{If_{q^{'}}},
\end{equation}
and {\it the q-Laplace transform between equivalence classes}
\begin{equation}
\label{ep6.25} L_{PR}:\Lambda_I\longrightarrow\boldsymbol{{\cal
U}_I},
\end{equation}
as
\begin{equation}
\label{ep6.26}
L_{PR}(\Lambda_{If_{q^{'}}})(p,q^{'})=L_{PR}(p,q^{'})=
L_R(f_{q^{'}})(p,q^{'}),
\end{equation}
one finds that $L_{PR}$ is one to one between equivalence classes
and is the analog of $F_{UTS}$ (the Umarov-Tsallis-Steinberg
q-Fourier transform \cite{umarov}) for the one to one q-Fourier
transform.

\setcounter{equation}{0}

\section{Examples}

We illustrate here with some examples the preceding developments.
As a first one we consider the q-Laplace transform of the
Heaviside's step function $f(x)=H(x)$. We have
\begin{equation}
\label{ep7.1} L(p,q)=[H(q-1)-H(q-2)]H[\Re(p)]
\int\limits_0^{\infty}[1+(q-1)px]^{\frac {1} {1-q}}]dx.
\end{equation}
Suitably manipulating (\ref{ep7.1}) leads to
\begin{equation}
\label{ep7.2} L(p,q)=[H(q-1)-H(q-2)]\frac {H[\Re(p)]} {(2-q)p}.
\end{equation}
For $f(x)=H(-x)$ we have
\begin{equation}
\label{ep7.3} L(p,q)=[H(q-1)-H(q-2)]H[-\Re(p)]
\int\limits_{-\infty}^0[1+(q-1)px]^{\frac {1} {1-q}}]dx,
\end{equation}
and, as a result,
\begin{equation}
\label{ep7.4} L(p,q)=[H(q-1)-H(q-2)]\frac {H[-\Re(p)]} {(2-q)p}.
\end{equation}
Taking now into account that $H(x)+H(-x)=1$ we get for the
q-Laplace transform of $f(x)=1$
\begin{equation}
\label{ep7.5} L(p,q)=[H(q-1)-H(q-2)]\frac {1} {(2-q)p}.
\end{equation} \vskip 4mm

\nd We evaluate now the transform of $f(x)=q^{'}H(x)$, i.e.,
\[L(p,q,q^{'})=[H(q-1)-H(q-2)]H[\Re(p)]\times\]
\begin{equation}
\label{ep7.6} \int\limits_0^{\infty}q^{'}
[1+(q-1)pxq^{'(q-1)}]^{\frac {1} {1-q}}]dx.
\end{equation}
One finds
\begin{equation}
\label{ep7.7} L(p,q,q^{'})=[H(q-1)-H(q-2)] \frac {q^{'(2-q)}}
{2-q} \frac {H[\Re(p)]} {p}.
\end{equation}
If we consider now the Laplace transform of the previous function
we face

\begin{equation}
\label{ep7.8} L_{PR}(p,q^{'})=[H(q^{'}-1)-H(q^{'}-2)] \frac
{q^{'(2-q^{'})}} {2-q^{'}} \frac {H[\Re(p)]} {p}.
\end{equation}

\vskip 4mm

\nd As a last example we evaluate the transform of the function
\begin{equation}
\label{ep7.9}
f(x)=
\begin{cases}
\left(\frac {\lambda} {x}\right)^{\beta}\;;\; x\in[a,b]\;;\; 0<a<b\;;\;\lambda>0 \\
0\;;\;x\; \rm{outside}\; [a,b].
\end{cases}
\end{equation}

\nd One has
\[L(p,q)=[H(q-1)-H(q-2)] H[\Re(p)]\times\]
\begin{equation}
\label{ep7.10} {\lambda}^{\beta} \int\limits_a^b
x^{-\beta}\{1-(1-q)p{\lambda}^{\beta(q-1)}
x^{1-\beta(q-1)}\}^{\frac {1} {1-q}}. \;dx
\end{equation}
Following the steps of a similar calculation made in \cite{tp4}
allow us to obtain
\[L(p,q)=[H(q-1)-H(q-2)]H[\Re(p)]\times\]
\[\left\{\left\{H(q-1)-H\left[q-\left(1+\frac {1} {\beta}\right)\right]\right\}\right.\times\]
\[\frac {(q-1){\lambda}^{\beta}} {(2-q)
[(q-1)p{\lambda}^{\beta}]^{\frac {1} {q-1}}}\times \]
\[\left\{a^{\frac {q-2} {q-1}}F\left(\frac {1} {q-1},\frac {2-q} {(q-1)[1-\beta(q-1)]},
\frac {1} {q-1} + \frac {\beta(2-q)} {1-\beta(q-1)};\right.\right.\]
\[\left.\frac {1} {(1-q)p{\lambda}^{\beta(q-1)}a^{1-\beta(q-1)}}\right)-\]
\[ b^{\frac {q-2} {q-1}}F\left(\frac {1} {q-1},\frac {2-q} {(q-1)[1-\beta(q-1)]},
\frac {1} {q-1} + \frac {\beta(2-q)} {1-\beta(q-1)};\right.\]
\[\left.\left.\frac {1} {(1-q)p{\lambda}^{\beta(q-1)}b^{1-\beta(q-1)}}\right)\right\}+\]
\[\left\{H\left[q-\left(1+\frac {1} {\beta}\right)\right]-H(q-2)\right\}
\frac {{\lambda}^{\beta}} {\beta-1}\times\]
\[\left\{a^{1-\beta}F\left(\frac {1} {q-1},\frac {\beta-1} {\beta(q-1)-1},
\frac {\beta q-2} {\beta(q-1)-1};\right.\right.\]
\[\left.(1-q)p{\lambda}^{\beta(q-1)}a^{1-\beta(q-1)}\right)-\]
\[b^{1-\beta}F\left(\frac {1} {q-1},\frac {\beta-1} {\beta(q-1)-1},
\frac {\beta q-2} {\beta(q-1)-1};\right.\]
\begin{equation}
\label{ep7.11}
\left.\left.\left.(1-q)p{\lambda}^{\beta(q-1)}b^{1-\beta(q-1)}\right)\right\}\right\}.
\end{equation}
and, if we take $\beta=1/(q-1)$, we get for out transform
\begin{equation}
\label{ep7.12} L_{PR}(p,q)=H[\Re(p)]\left[H(q-1)-H(q-2)\right]
\left[1-(1-q)p\lambda\right]^{\frac {1} {1-q}}.
\end{equation}

\setcounter{equation}{0}

\section{Series expansion of the q-Laplace transform}

Consider the function
\[\{1-(1-q)px[f(x)]^{q-1}\}^{\frac {1} {1-q}}.\]
Using the series expansions of the logarithm and the exponential
function, we can write
\[\{1-(1-q)px[f(x)]^{q-1}\}^{\frac {1} {1-q}}=
e^{\frac {1} {1-q}\ln\{1-(1-q)px[f(x)]^{q-1}\}}=\]
\[e^{\frac {1} {1-q}\sum\limits_{n=1}^{\infty}
\frac {(-1)^{n+1}} {n}(q-1)^n(px)^n[f(x)]^{n(q-1)}}=\]
\[e^{\frac {1} {1-q}\sum\limits_{n=1}^{\infty}
\frac {(-1)^{n+1}} {n}(q-1)^n(px)^n
e^{n(q-1)\ln f(x)}}=\]
\[e^{\frac {1} {1-q}\sum\limits_{n=1}^{\infty}
\frac {(-1)^{n+1}} {n}(q-1)^n(px)^n
\sum\limits_{m=0}^{\infty}\frac {n^m} {m!}
(q-1)^m [\ln f(x)]^m}=\]
\[e^{\left[\sum\limits_{n=1}^{\infty}\sum\limits_{m=0}^{\infty}
\frac {(-1)^n n^{m-1}} {m!} (px)^n \ln^m[f(x)](q-1)^{n+m-1}\right]}=\]
\begin{equation}
\label{ep1.1}
e^{\left\{\sum\limits_{n=0}^{\infty}\left[\sum\limits_{m=0}^{n}
\frac {(-1)^{n+1-m}(n+1-m)^{m-1}} {m!} (px)^{n-m+1}
 \ln^m[f(x)]\right](q-1)^n\right\}}.
\end{equation}
Let $g(x,p,n)$ be given by
\begin{equation}
\label{ep1.2} g(x,k,n)=\sum\limits_{m=0}^n \frac
{(-1)^{n+1-m}(n-m+1)^{m-1}} {m!}(px)^{n-m+1}\ln^m[f(x)].
\end{equation}
Then,
\begin{equation}
\label{ep1.3}
\{1-(1-q)px[f(x)]^{q-1}\}=h(x,p,q)=e^{\;\sum\limits_{n=0}^{\infty}
g(x,p,n)(q-1)^n},
\end{equation}
or
\begin{equation}
\label{ep1.4} h(x,p,q)=e^{-px}e^{\;\sum\limits_{n=1}^{\infty}
g(x,p,n)(q-1)^n}.
\end{equation}
Minding  the expansion of the exponential function we have
\begin{equation}
\label{ep1.5} e^{\;\sum\limits_{n=1}^{\infty}g(x,p,n)(q-1)^n}=
\sum\limits_{l=0}^{\infty}\frac {\left(\sum\limits_{n=1}^{\infty}
g(x,p,n)(q-1)^n\right)^l} {l!},
\end{equation}
and, as a consequence,
\begin{equation}
\label{ep1.6} h(x,p,q)=e^{ikx}\left[1+\sum\limits_{n=1}^{\infty},
l(x,p,n)\right]
\end{equation}
where
\[l(x,p,n)=\frac {1} {n!}\sum\limits_{s=n}^{\infty}
\sum\limits_{s_1=1}^{s-n+1}\sum\limits_{s_2=1}^{s-s_1-n+2}
\cdot\cdot\cdot\sum\limits_{s_{n-1}=1}^{s-s_1-s_2-\cdot\cdot\cdot
-s_{n-2}-1}\]
\[g(x,p,s_1)g(x,p,s_2)\cdot\cdot\cdot
g(x,p,s_{n-1})\times\]
\begin{equation}
\label{ep1.7} g(x,p,s-s_1-s_2-\cdot\cdot\cdot-s_{n-1})(q-1)^s.
\end{equation}
One can write the q-Laplace transform as
\[L(p,q)=[H(q-1)-H(q-2)]\times \]
\begin{equation}
\label{ep1.8}
\left\{H[\Re(p)]\int\limits_0^{\infty} f(x)h(x,p,q)
\;dx -
H[-\Re(p)]\int\limits_{-\infty}^0 f(x)
h(x,p,q) \;dx\right\}.
\end{equation}

\setcounter{equation}{0}

\section{The q-Laplace transform of the q-Gaussian}

\nd Our purpose is to calculate the q-Laplace transform of the
q-Gaussian. As this becomes a too complex task in the general
case, we content ourselves with a first-order expansion  in powers
of $q-1$. Accordingly,
\begin{equation}
\label{ep4.1} h(x,p,q)=e^{-px} [1+g(x,p,1)(q-1)],
\end{equation}
with
\begin{equation}
\label{ep4.2} g(x,p,1)=\frac {(px)^2} {2} - px \ln[f(x)].
\end{equation}
Then, up  to first order we have for the q-Laplace transform
\[L(p,q)=[H(q-1)-H(q-2)]\times \]
\[\left\{H[\Re(p)]\int\limits_0^{\infty} \left\{1+
\left\{\frac {(px)^2} {2} - px \ln[f(x)]\right\}(q-1)\right\}f(x)e^{-px}
\;dx -\right.\]
\begin{equation}
\label{ep4.3} \left.H[-\Re(p)]\int\limits_{-\infty}^0
\left\{1+\left\{\frac {(px)^2} {2} - px
\ln[f(x)]\right\}(q-1)\right\}f(x)e^{-px} \;dx\right\}.
\end{equation}
Let $G(k)$ and $G(k,\beta)$ be given by
\begin{equation}
\label{ep4.4} G(p)=\left\{H[\Re(p)]\int\limits_0^{\infty}
f(x)e^{-px}\;dx -
H[-\Re(p)]\int\limits_{-\infty}^0f(x)e^{-px}\;dx\right\},
\end{equation}
\begin{equation}
\label{ep4.5} G(p,\beta)=\left\{H[\Re(p)]\int\limits_0^{\infty}
[f(x)]^{\beta}e^{-px}\;dx -
H[-\Re(p)]\int\limits_{-\infty}^0[f(x)]^{\beta}
e^{-px}\;dx\right\}.
\end{equation}
We write the q-Laplace transform in the form
\[L(p,q)=[H(q-1)-H(q-2)]\times \]
\begin{equation}
\label{ep4.6} G(p)+\left[\frac {p^2} {2}\frac {{\partial}^2}
{\partial p^2}G(p)+p\frac {\partial} {\partial p} \frac {\partial}
{\partial\beta}G(p,\beta)\right]_{\beta=1} (q-1).
\end{equation}
Let $f(x)$ be the q-Gaussian
\begin{equation}
\label{ep4.7} f(x)=C_{q^{'}}[1+ (q^{'}-1)\alpha x^2]^{\frac {1}
{1-q^{'}}},
\end{equation}
where
\begin{equation}
\label{ep4.8} C_{q^{'}}=\frac {\sqrt{(q^{'}-1)\alpha}}
{B\left(\frac {1} {2}, \frac {1} {q^{'}-1}\frac {1}
{2}\right)}\;\;\;q^{'}\neq 1,
\end{equation}
\begin{equation}
\label{ep4.9} C_1=\sqrt{\frac {\alpha} {\pi}}.
\end{equation}
Using results of \cite{tt3} we obtain
\[G(p,q^{'})=H[\Re(p)]C_{q^{'}}\frac {\sqrt{\pi}} {2}
\frac {\Gamma\left(\frac {2-q^{'}} {1-q^{'}}\right)}
{[(q^{'}-1)\alpha]^{\frac {1} {1-q^{'}}}}
\left[\frac {2} {(q^{'}-1)\alpha p }\right]^{\frac {2-q^{'}}
{1-q^{'}}-\frac {1} {2}}\times\]
\[\left\{{\bf H}_{\frac {2-q^{'}} {1-q^{'}}-\frac {1} {2}}
\left(\frac {p} {(q^{'}-1)\alpha}\right)-
{\bf N}_{\frac {2-q^{'}} {1-q^{'}}-\frac {1} {2}}
\left(\frac {p} {(q^{'}-1)\alpha}\right)\right\}-\]
\[H[-\Re(p)]C_{q^{'}}\frac {\sqrt{\pi}} {2}
\frac {\Gamma\left(\frac {2-q^{'}} {1-q^{'}}\right)}
{[(q^{'}-1)\alpha]^{\frac {1} {1-q^{'}}}}
\left[\frac {2} {(1-q^{'})\alpha p }\right]^{\frac {2-q^{'}}
{1-q^{'}}-\frac {1} {2}}\times\]
\begin{equation}
\label{ep4.10} \left\{{\bf H}_{\frac {2-q^{'}} {1-q^{'}}-\frac {1}
{2}} \left(\frac {p} {(1-q^{'})\alpha}\right)- {\bf N}_{\frac
{2-q^{'}} {1-q^{'}}-\frac {1} {2}} \left(\frac {p}
{(1-q^{'})\alpha}\right)\right\},
\end{equation}
and
\[G(p,q^{'},\beta)=H[\Re(p)]C_{q^{'}}^{\beta}\frac {\sqrt{\pi}} {2}
\frac {\Gamma\left(\frac {\beta + 1-q^{'}} {1-q^{'}}\right)}
{[(q^{'}-1)\alpha]^{\frac {\beta} {1-q^{'}}}}
\left[\frac {2} {(q^{'}-1)\alpha p }\right]^{\frac {\beta +1-q^{'}}
{1-q^{'}}-\frac {1} {2}}\times\]
\[\left\{{\bf H}_{\frac {\beta+1-q^{'}} {1-q^{'}}-\frac {1} {2}}
\left(\frac {p} {(q^{'}-1)\alpha}\right)-
{\bf N}_{\frac {\beta+1-q^{'}} {1-q^{'}}-\frac {1} {2}}
\left(\frac {p} {(q^{'}-1)\alpha}\right)\right\}-\]
\[H[-\Re(p)]C_{q^{'}}^{\beta}\frac {\sqrt{\pi}} {2}
\frac {\Gamma\left(\frac {\beta +1-q^{'}} {1-q^{'}}\right)}
{[(q^{'}-1)\alpha]^{\frac {\beta} {1-q^{'}}}}
\left[\frac {2} {(1-q^{'})\alpha p }\right]^{\frac {\beta +1-q^{'}}
{1-q^{'}}-\frac {1} {2}}\times\]
\begin{equation}
\label{ep4.11} \left\{{\bf H}_{\frac {\beta+1-q^{'}}
{1-q^{'}}-\frac {1} {2}} \left(\frac {p} {(1-q^{'})\alpha}\right)-
{\bf N}_{\frac {\beta+1-q^{'}} {1-q^{'}}-\frac {1} {2}}
\left(\frac {p} {(1-q^{'})\alpha}\right)\right\},
\end{equation}
where ${\bf H}$ and ${\bf N}$ are the Struve and Neumann
functions, respectively. The q-Laplace transform of the q-Gaussian
is now
\[L(p,q,q^{'})=[H(q-1)-H(q-2)]\times \]
\begin{equation}
\label{ep4.12} G(p,q^{'})+\left[\frac {p^2} {2}\frac
{{\partial}^2} {\partial p^2}G(p,q^{'})+k\frac {\partial}
{\partial p} \frac {\partial}
{\partial\beta}G(p,q^{'},\beta)\right]_{\beta=1} (q-1).
\end{equation}
If instead of using the bilateral q-Laplace transform we use the
unilateral one, we should replace (\ref{ep4.10}) and
(\ref{ep4.11}), respectively, by
\[G(p,q^{'})=H[\Re(p)]C_{q^{'}}\frac {\sqrt{\pi}} {2}
\frac {\Gamma\left(\frac {2-q^{'}} {1-q^{'}}\right)}
{[(q^{'}-1)\alpha]^{\frac {1} {1-q^{'}}}}
\left[\frac {2} {(q^{'}-1)\alpha p }\right]^{\frac {2-q^{'}}
{1-q^{'}}-\frac {1} {2}}\times\]
\begin{equation}
\label{ep4.13} \left\{{\bf H}_{\frac {2-q^{'}} {1-q^{'}}-\frac {1}
{2}} \left(\frac {p} {(q^{'}-1)\alpha}\right)- {\bf N}_{\frac
{2-q^{'}} {1-q^{'}}-\frac {1} {2}} \left(\frac {p}
{(q^{'}-1)\alpha}\right)\right\},
\end{equation}
and
\[G(p,q^{'},\beta)=H[\Re(p)]C_{q^{'}}^{\beta}\frac {\sqrt{\pi}} {2}
\frac {\Gamma\left(\frac {\beta + 1-q^{'}} {1-q^{'}}\right)}
{[(q^{'}-1)\alpha]^{\frac {\beta} {1-q^{'}}}}
\left[\frac {2} {(q^{'}-1)\alpha p }\right]^{\frac {\beta +1-q^{'}}
{1-q^{'}}-\frac {1} {2}}\times\]
\begin{equation}
\label{ep4.14} \left\{{\bf H}_{\frac {\beta+1-q^{'}}
{1-q^{'}}-\frac {1} {2}} \left(\frac {p} {(q^{'}-1)\alpha}\right)-
{\bf N}_{\frac {\beta+1-q^{'}} {1-q^{'}}-\frac {1} {2}}
\left(\frac {p} {(q^{'}-1)\alpha}\right)\right\}.
\end{equation}

\setcounter{equation}{0}

\section{The q-Laplace transform of the q-Gaussian for  fixed q}

\nd In this section we deal with the q-Laplace transform of the
q-Gaussian for  fixed $q$. We have
\[L_{PR}(p,q)=H[\Re(p)]\int\limits_0^{\infty} C_q[1+(q-1)\alpha x^2]^{
\frac {1} {1-q}}\times\]
\[\left\{1+(1-q)ikx\left\{C_q[1+(q-1)\alpha x^2]^{\frac {1} {1-q}}
\right\}({q-1}\right\}^{\frac {1} {1-q}}\;dx-\]
\[H[-\Re(p)]\int\limits_{-\infty}^0 C_q[1+(q-1)\alpha x^2]^{
\frac {1} {1-q}}\times\]
\begin{equation}
\label{ep5.1} \left\{1+(q-1)px\left\{C_q[1+(q-1)\alpha x^2]^{\frac
{1} {1-q}} \right\}({q-1}\right\}^{\frac {1} {1-q}}\;dx,
\end{equation}
$1\leq q <2$. Simplifying terms we obtain
\[L_{PR}(p,q)=H[\Re(p)]\int\limits_0^{\infty} C_q
[(q-1)\alpha x^2+(q-1)C_q^{q-1}px+1
]^{\frac {1} {1-q}}\;dx-\]
\begin{equation}
\label{ep5.2} H[-\Re(p)]\int\limits_{-\infty}^0 C_q [(q-1)\alpha
x^2+e^{\frac {i\pi} {2}}(q-1)C_q^{q-1}kx+1 ]^{\frac {1}
{1-q}}\;dx.
\end{equation}
Effecting the change of variables $\sqrt{(q-1)\alpha}\;x=y$, the
q-Laplace transform becomes
\[L_{PR}(p,q)=\frac {H[\Re(p)]} {\sqrt{(q-1)\alpha}}
\int\limits_0^{\infty} C_q
\left[y^2+C_q^{q-1}\sqrt{\frac {q-1} {\alpha}}py+1
\right]^{\frac {1} {1-q}}\;dy-\]
\begin{equation}
\label{ep5.3} \frac {H[-\Re(p)]} {\sqrt{(q-1)\alpha}}
\int\limits_0^{\infty} C_q \left[y^2-C_q^{q-1}\sqrt{\frac {q-1}
{\alpha}}py+1 \right]^{\frac {1} {1-q}}\;dy.
\end{equation}
Using  results given in \cite{tt5} we see that  (${\bf P}_\nu^\mu$
is the associated Legendre function)
\[{\bf P}_\nu^\mu(z)=\frac {2^\mu \Gamma(1-2\mu)
(z^2-1)^{\frac {\mu} {2}}} {\Gamma(1-\mu)\Gamma(-\mu-\nu)
\Gamma(\nu-\mu+1)}\times\]
\begin{equation}
\label{ep5.4}
\int\limits_0^\infty\left(1+2tz+t^2\right)^{\mu-\frac {1} {2}}
t^{-1-\nu-\mu}\;dt,
\end{equation}
and we can write
\begin{equation}
\label{ep5.5}
\int\limits_0^\infty\left(1+2tz+t^2\right)^{\mu-\frac {1} {2}}
t^{-1-\nu-\mu}\;dt=\Gamma(-\mu) 2^{-\mu-1} (z^2-1)^{\frac {\mu}
{2}}{\bf P}_{-\mu-1}^\mu(z),
\end{equation}
Let $\gamma , \mu$ be given by:
\begin{equation}
\label{ep5.6} \gamma=\frac {C_q^{q-1}} {2} \sqrt{\frac {q-1}
{\alpha}} \;\;\;\; \mu=\frac {1} {1-q}+\frac {1} {2},
\end{equation}
so that
\[L_{PR}(p,q)=C_q\frac {\Gamma(-\mu)} {\sqrt{(q-1)\alpha}}
2^{-\mu-1} (\gamma^2 p^2 - 1)^{\frac {
\mu} {2}}\times\]
\begin{equation}
\label{ep5.7} \left\{H[\Re(p)]{\bf P}_{-1-\mu}^{\mu}(\gamma p)
-H[-\Re(p)] {\bf P}_{-1-\mu}^{\mu}(-\gamma p)\right\},
\end{equation}
which is the q-Laplace Transform of the q-Gaussian (on the complex
plane) for fixed $q$. If instead of using the bilateral q-Laplace
transform we use the unilateral one, we just obtain
\begin{equation}
\label{ep5.8} L_{PR}(p,q)=C_q\frac {\Gamma(-\mu)}
{\sqrt{(q-1)\alpha}} 2^{-\mu-1} (\gamma^2 p^2 - 1)^{\frac { \mu}
{2}} H[\Re(p)]{\bf P}_{-1-\mu}^{\mu}(\gamma p).
\end{equation}

\setcounter{equation}{0}

\section{The q-partition function}

As it is well known, the partition function of a system is the
unilateral Laplace transform of the density of states \cite{chem}.
Thus, the q-partition function should be  defined as the q-Laplace
transform of the density of states  $f(u)$, in which the complex
variable is now ${\cal B}$  (in place of $p$). We have
\[Z({\cal B},q)=[H(q-1)-H(q-2)]\times \]
\begin{equation}
\label{ep8.1} H[\Re({\cal B})]\int\limits_0^{\infty}
f(u)\{1-(1-q){\cal B}u[f(u)]^{(q-1)}\}^{\frac {1} {1-q}}\;du.
\end{equation}
If the density of states depends on $q$ we can define it following
the  $L_{PR}$ definition given in section 2.
\[Z_{PR}({\cal B},q)=[H(q-1)-H(q-2)]\times \]
\begin{equation}
\label{ep8.2} H[\Re({\cal B})]\int\limits_0^{\infty}
f_q(u)\{1-(1-q){\cal B}u[f_q(u)]^{(q-1)}\}^{\frac {1} {1-q}}\;du.
\end{equation}
For example,  if the density of states is a q-exponential
\begin{equation}
\label{ep8.3} f_q(u)=[1+(q-1)\alpha u]^{\frac {1} {1-q}},
\end{equation}
where $\alpha>0$, the q-partition function is
\begin{equation}
\label{ep8.4} Z_{PR}({\cal B},q)=[H(q-1)-H(q-2)] \frac
{H[\Re({\cal B})]} {{\cal B}+\alpha}.
\end{equation}
\nd An  important case is $f(u)= constant=H(x)$. We face a
non-degenerate energy spectrum (the one dimensional harmonic
oscillator, for instance). We have then,
\begin{equation}
\label{ep8.5} Z({\cal B},q)=[H(q-1)-H(q-2)] \frac {H[\Re({\cal
B})]} {(2-q){\cal B}}.
\end{equation}

\section*{Conclusions}

\nd We have here developed the q-Laplace transform, thus
incorporating it to the Tsallis' arsenal. We have studied its main
properties and analyzed some instructive examples. The
particularly important case of the q-Gaussian has been discussed
in some detail. Finally, we have derived the q-partition function
from the q-Laplace transform. As is also the case with the
q-Fourier transform, we realize that the q-Laplace transform is
essentially a transformation between equivalence classes.

\newpage

\newpage

\section{Appendix: Tempered ultradistributions and
distributions of exponential type }

\setcounter{equation}{0}

\nd Many readers will surely benefit from a brief summary of the
main properties of distributions of exponential type and of
tempered ultradistributions.

\nd {\bf Notations}. The notations are almost textually taken from
Ref. \cite{tp2}. Let $\boldsymbol{{\mathbb{R}}^n}$ (res.
$\boldsymbol{{\mathbb{C}}^n}$) be the real (resp. complex)
n-dimensional space whose points are denoted by
$x=(x_1,x_2,...,x_n)$ (resp $z=(z_1,z_2,...,z_n)$). We shall use
the notations:

(a) $x+y=(x_1+y_1,x_2+y_2,...,x_n+y_n)$\; ; \;
    $\alpha x=(\alpha x_1,\alpha x_2,...,\alpha x_n)$

(b)$x\geqq 0$ means $x_1\geqq 0, x_2\geqq 0,...,x_n\geqq 0$

(c)$x\cdot y=\sum\limits_{j=1}^n x_j y_j$

(d)$\mid x\mid =\sum\limits_{j=1}^n \mid x_j\mid$

\nd Let $\boldsymbol{{\mathbb{N}}^n}$ be the set of n-tuples of
natural numbers. If $p\in\boldsymbol{{\mathbb{N}}^n}$, then
$p=(p_1, p_2,...,p_n)$, and $p_j$ is a natural number, $1\leqq
j\leqq n$. $p+q$ stands for $(p_1+q_1, p_2+q_2,..., p_n+q_n)$ and
$p\geqq q$ means $p_1\geqq q_1, p_2\geqq q_2,...,p_n\geqq q_n$.
$x^p$ entails $x_1^{p_1}x_2^{p_2}... x_n^{p_n}$. We shall denote
by $\mid p\mid=\sum\limits_{j=1}^n p_j $ and call $D^p$  the
differential operator
${\partial}^{p_1+p_2+...+p_n}/\partial{x_1}^{p_1}
\partial{x_2}^{p_2}...\partial{x_n}^{p_n}$

\nd For any natural $k$ we define $x^k=x_1^k x_2^k...x_n^k$ and
${\partial}^k/\partial x^k= {\partial}^{nk}/\partial x_1^k\partial
x_2^k...\partial x_n^k$

\nd The space $\boldsymbol{{\cal H}}$  of test functions such that
$e^{p|x|}|D^q\phi(x)|$ is bounded for any $p$ and $q$, being
defined [see Ref. (\cite{tp2})] by means of the countably set of
norms
\begin{equation}
\label{ep2.1} {\|\hat{\phi}\|}_p=\sup_{0\leq q\leq p,\,x} e^{p|x|}
\left|D^q \hat{\phi} (x)\right|\;\;\;,\;\;\;p=0,1,2,...
\end{equation}

\nd The space of continuous linear functionals defined on
$\boldsymbol{{\cal H}}$ is the space
$\boldsymbol{{\Lambda}_{\infty}}$ of the distributions of the
exponential type given by ( ref.\cite{tp2} ).
\begin{equation}
\label{ep2.2} T=\frac {{\partial}^k} {\partial x^k} \left[
e^{k|x|}f(x)\right]
\end{equation}
where $k$ is an integer such that $k\geqq 0$ and $f(x)$ is a
bounded continuous function. \nd In addition we have
$\boldsymbol{{\cal H}}\subset\boldsymbol{{\cal S}}
\subset\boldsymbol{{\cal S}^{'}}\subset
\boldsymbol{{\Lambda}_{\infty}}$, where $\boldsymbol{{\cal S}}$ is
the Schwartz space of rapidly decreasing test functions
(ref\cite{tp6}).

The Fourier transform of a function $\hat{\phi}\in
\boldsymbol{{\cal H}}$ is
\begin{equation}
\label{ep3.1} \phi(z)=\frac {1} {2\pi}
\int\limits_{-\infty}^{\infty}\overline{\hat{\phi}}(x)\;e^{iz\cdot
x}\;dx
\end{equation}
According to ref.\cite{tp2}, $\phi(z)$ is entire analytic and
rapidly decreasing on straight lines parallel to the real axis. We
shall call $\boldsymbol{{\cal H}}$ the set of all such functions.
\begin{equation}
\label{ep3.2} \boldsymbol{{\cal H}}={\cal
F}\left\{\boldsymbol{{\cal H}}\right\}
\end{equation}
The topology in $\boldsymbol{{\cal H}}$ is defined by the
countable set of semi-norms:
\begin{equation}
\label{ep3.4} {\|\phi\|}_{k} = \sup_{z\in V_k} |z|^k|\phi (z)|,
\end{equation}
where $V_k=\{z=(z_1,z_2,...,z_n)\in\boldsymbol{{\mathbb{C}}^n}:
\mid Im z_j\mid\leqq k, 1\leqq j \leqq n\}$

\nd The dual of $\boldsymbol{{\cal H}}$ is the space
$\boldsymbol{{\cal U}}$ of tempered ultradistributions [see Ref.
(\cite{tp2} )]. In other words, a tempered ultradistribution is a
continuous linear functional defined on the space
$\boldsymbol{{\cal H}}$ of entire functions rapidly decreasing on
straight lines parallel to the real axis. \nd Moreover, we have
$\boldsymbol{{\cal H}}\subset\boldsymbol{{\cal S}}
\subset\boldsymbol{{\cal S}^{'}}\subset \boldsymbol{{\cal U}}$.

$\boldsymbol{{\cal U}}$ can also be characterized in the following
way [see Ref. (\cite{tp2} )]: let $\boldsymbol{{\cal A}_{\omega}}$
be the space of all functions $F(z)$ such that:

${\Large {\boldsymbol{A)}}}$- $F(z)$ is analytic for $\{z\in
\boldsymbol{{\mathbb{C}}^n} : |Im(z_1)|>p,
|Im(z_2)|>p,...,|Im(z_n)|>p\}$.

${\Large {\boldsymbol{B)}}}$- $F(z)/z^p$ is bounded continuous  in
$\{z\in \boldsymbol{{\mathbb{C}}^n} :|Im(z_1)|\geqq
p,|Im(z_2)|\geqq p, ...,|Im(z_n)|\geqq p\}$, where $p=0,1,2,...$
depends on $F(z)$.

\nd Let $\boldsymbol{\Pi}$ be the set of all $z$-dependent
pseudo-polynomials, $z\in \boldsymbol{{\mathbb{C}}^n}$. Then
$\boldsymbol{{\cal U}}$ is the quotient space

${\Large {\boldsymbol{C)}}}$- $\boldsymbol{{\cal
U}}=\boldsymbol{{\cal A}_{\omega}/\Pi}$

\nd By a pseudo-polynomial we understand a function of $z$ of the
form $\;\;$ $\sum_s z_j^s G(z_1,...,z_{j-1},z_{j+1},...,z_n)$ with
$G(z_1,...,z_{j-1},z_{j+1},...,z_n)\in\boldsymbol{{\cal
A}_{\omega}}$

\nd Due to these properties it is possible to represent any
ultradistribution as [see Ref. (\cite{tp2} )]
\begin{equation}
\label{ep3.6} F(\phi)=<F(z), \phi(z)>=\oint\limits_{\Gamma} F(z)
\phi(z)\;dz
\end{equation}
$\Gamma={\Gamma}_1\cup{\Gamma}_2\cup ...{\Gamma}_n,$  where the
path ${\Gamma}_j$ runs parallel to the real axis from $-\infty$ to
$\infty$ for $Im(z_j)>\zeta$, $\zeta>p$ and back from $\infty$ to
$-\infty$ for $Im(z_j)<-\zeta$, $-\zeta<-p$. ($\Gamma$ surrounds
all the singularities of $F(z)$).

\nd Eq. (\ref{ep3.6}) will be our fundamental representation for a
tempered ultradistribution. Sometimes use will be made of the
``Dirac formula" for ultradistributions [see Ref. (\cite{tp1})]
\begin{equation}
\label{ep3.7} F(z)=\frac {1} {(2\pi
i)^n}\int\limits_{-\infty}^{\infty} \frac {f(t)}
{(t_1-z_1)(t_2-z_2)...(t_n-z_n)}\;dt
\end{equation}
where the ``density'' $f(t)$ is such that
\begin{equation}
\label{ep3.8} \oint\limits_{\Gamma} F(z) \phi(z)\;dz =
\int\limits_{-\infty}^{\infty} f(t) \phi(t)\;dt.
\end{equation}
While $F(z)$ is analytic on $\Gamma$, the density $f(t)$ is in
general singular, so that the r.h.s. of (\ref{ep3.8}) should be
interpreted in the sense of distribution theory.

\nd Another important property of the analytic representation is
the fact that on $\Gamma$, $F(z)$ is bounded by a power of $z$
\cite{tp2}
\begin{equation}
\label{ep3.9} |F(z)|\leq C|z|^p,
\end{equation}
where $C$ and $p$ depend on $F$.

\nd The representation (\ref{ep3.6}) implies that the addition of
a pseudo-polynomial $P(z)$ to $F(z)$ does not alter the
ultradistribution:
\[\oint\limits_{\Gamma}\{F(z)+P(z)\}\phi(z)\;dz=
\oint\limits_{\Gamma} F(z)\phi(z)\;dz+\oint\limits_{\Gamma}
P(z)\phi(z)\;dz\] However,
\[\oint\limits_{\Gamma} P(z)\phi(z)\;dz=0.\]
As $P(z)\phi(z)$ is entire analytic in some of the variables $z_j$
(and rapidly decreasing), we obtain:
\begin{equation}
\label{ep3.10} \oint\limits_{\Gamma} \{F(z)+P(z)\}\phi(z)\;dz=
\oint\limits_{\Gamma} F(z)\phi(z)\;dz.
\end{equation}

\end{document}